\documentclass[10pt,twocolumn,a4paper]{article}
\usepackage[margin=16mm,columnsep=6mm]{geometry}
\usepackage{newtxtext,newtxmath}
\usepackage{graphicx}
\usepackage{booktabs}
\usepackage{multirow}
\usepackage[font=small,labelfont=bf,skip=3pt]{caption}
\usepackage[hidelinks]{hyperref}
\usepackage{enumitem}
\setlist{nosep,leftmargin=*}
\makeatletter
\renewcommand{\section}{\@startsection{section}{1}{\z@}{1.4ex plus .5ex}{0.8ex}{\normalfont\large\bfseries}}
\renewcommand{\subsection}{\@startsection{subsection}{2}{\z@}{1.0ex plus .3ex}{0.4ex}{\normalfont\normalsize\bfseries}}
\makeatother

\title{\vspace{-8mm}\Large\bfseries Why Do Low-Activity Speakers Disappear in Speaker Diarization?\\[1mm]\large Decomposing Lost Speakers and What Training-Free Fixes Can Recover\vspace{-2mm}}
\author{Fumiaki Yamaguchi\\{\small Independent Researcher\quad\texttt{panchorange2203@gmail.com}}}
\date{\vspace{-6mm}}

\begin{document}
\maketitle

\begin{abstract}
Speaker diarization can drop people who speak only a little from its output entirely, and such losses barely affect the diarization error rate (DER), which measures the fraction of misattributed time.
We examine Nemotron-3, an end-to-end diarization model with eight output slots, on 990 recordings from four corpora, and count lost speakers by how they were lost (overlap, merge, miss).
Nine in ten lost speakers are low-activity speakers with less than 10\,s of total speech, and the cause splits by the number of speakers in the recording.
In recordings with nine or more speakers, merges and misses caused by running out of slots dominate; re-running the model on short windows only for recordings that fill all eight slots, without any training, reduces these losses from 228 to 101 speakers, while extra output speakers grow from 5 to 50.
In recordings with up to eight speakers, 60\% of the losses are speakers hidden under someone else's speech, and none of the training-free methods we tried reduced them.
\end{abstract}

\section{Introduction}
\begin{figure*}[t]
\centering
\includegraphics[width=0.52\textwidth]{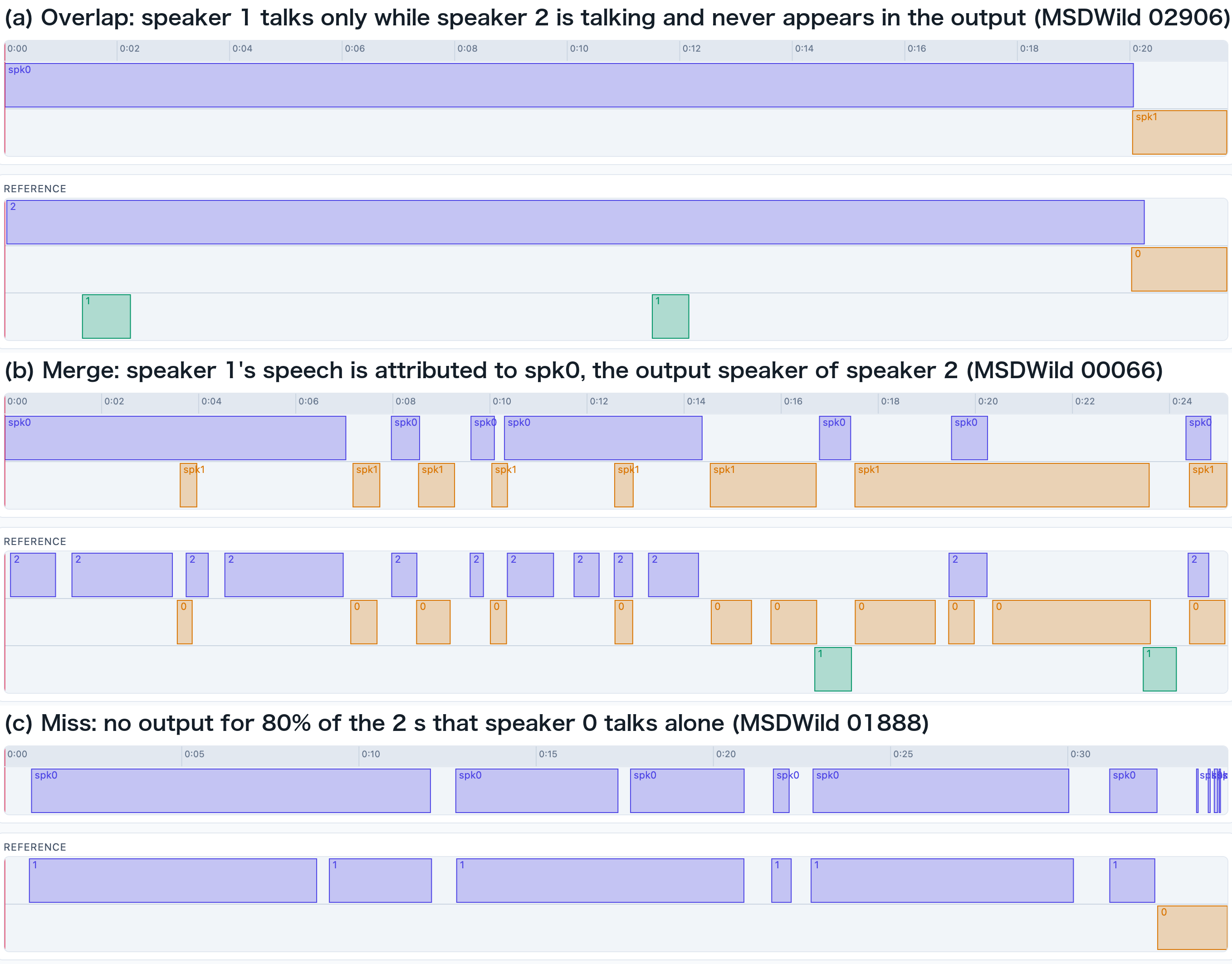}
\caption{Examples of the three loss types (voxmap studio). Each panel shows the Nemotron-3 output (top) and the reference (bottom); speakers matched by the optimal mapping share a color.
In every example the lost speaker (green or orange) has recall 0.}
\label{fig:types}
\end{figure*}

Speaker diarization estimates who spoke when, e.g., for meeting minutes and subtitles. If someone who only gives backchannels or short questions is dropped, everything they said vanishes from the transcript, yet DER, which divides misattributed time by total speech time, barely changes.
In the recording of Fig.~\ref{fig:types}(a), one of three speakers is lost, yet DER is 7.6\%, and 0.0\% with a 0.25\,s collar and overlap excluded.

End-to-end diarizers such as Sortformer~\cite{sortformer,streaming-sortformer} output a speech probability for a fixed number of speakers, so the number of speakers they can output (slots) is capped.
We study Nemotron-3, which has eight slots and a lower DER than a pyannote-based pipeline in our evaluation (Table~\ref{tab:main}), decompose what it means for a speaker to be lost, and ask whether each component can be reduced without training.
Our contributions are:
\begin{itemize}
\item We classify each lost speaker by how it was lost and show that lost speakers are those who speak briefly, and that the cause splits by the number of speakers into \emph{slot shortage} and \emph{overlap}.
\item We show that losses from slot shortage can be halved, with recordings of up to eight speakers left almost unchanged, by a simple training-free method that leaves the trained model untouched and re-runs it on windows only for recordings that fill all eight slots (the window-and-link idea itself follows prior work).
\item We show that losses in recordings with up to eight speakers are not reduced by the training-free methods we tried, and that counting lost speakers alone can overstate improvements.
\end{itemize}

\section{Related Work}
Besides DER, diarization is evaluated with JER~\cite{dihard}, which averages errors per speaker, BER~\cite{ber}, which emphasizes short segments and low-activity speakers, and CDER~\cite{cder}, which counts utterances.
Samarakoon and Zhu~\cite{beyond-der} showed that EEND under-counts speakers in recordings with five or more speakers, that DER hides this, and retrained the model with a speaker-weighted loss. We instead classify each lost speaker by how it was lost and measure how far training-free methods can go.
The speaker cap can be lifted by iterative inference that removes the training-time maximum (EEND-EDA~\cite{eendeda}) or by running inference per window and linking window speakers. EEND-VC~\cite{eendvc} clusters window-level speaker embeddings, and EEND-GLA~\cite{eendgla} combines global (whole-recording) and local (window) attractors (DiariZen~\cite{diarizen} is of the same family).
Chen et al.~\cite{chen-sortformer} split recordings into windows for Sortformer without training and cluster per-window mean embeddings to handle long recordings. Our windowing follows the same idea; the difference is that we choose which recordings to window by whether all slots are used, not by duration.
For overlaps we also try PixIT~\cite{pixit}, which jointly trains diarization and separation.

\section{Lost Speakers and Their Decomposition}
\label{sec:def}
Reference and output speakers are matched with the same optimal one-to-one mapping as DER.
A reference speaker's \emph{recall} is the fraction of their speech covered by the matched output speaker.
A reference speaker with recall 0 is \emph{lost}, one with recall $\geq 0.5$ is \emph{detected}, and one with less than 10\,s of total speech is a \emph{low-activity speaker}.
An output speaker matched to no reference speaker is an \emph{extra speaker}; this is a count of speakers, distinct from DER false alarm (seconds of output where nobody speaks).

Speakers with recall below 0.5 get a \emph{type} in this order: \emph{overlap} if at least half of their speech overlaps others, \emph{miss} if nothing is output for at least half of their solo speech, \emph{merge} if at least half of their solo speech is attributed to one other output speaker, and \emph{split} otherwise.
Overlap is tested first because solo-speech tests are unreliable for speakers who mostly overlap; this inflates the overlap share accordingly.
Being lost is the outcome and the type its breakdown; a miss is one kind of loss (Fig.~\ref{fig:types}).

Counting lost speakers alone favors methods that output speakers longer, which hit hidden speech by chance (Sec.~\ref{sec:overlap}), so we always report lost, detected and extra speakers together.
Whether extra speakers increase over the baseline is tested with a one-sided Wilcoxon signed-rank test on per-recording differences; recordings with zero difference drop out of the test, so we also report how many recordings changed.

\section{Experimental Setup}
\label{sec:setup}
\textbf{Data.}
We use 990 recordings: ICSI~\cite{icsi} (75), AMI~\cite{ami} (16, test), VoxConverse~\cite{voxconverse} (232, test), and the many (177) and few (490) subsets of MSDWild~\cite{msdwild}.
They are split 3:7 by the MD5 of the file name into calib (309) and eval (681).
Thresholds are chosen on calib and methods are compared on eval (except the 30-minute condition in Sec.~\ref{sec:method}, which was also chosen after looking at eval results of the adaptive variant).
The loss analysis (Sec.~\ref{sec:analysis}) chooses no thresholds and uses all 990 recordings.
Every number is reported with the set in Table~\ref{tab:sets} it was counted on.
Unless noted, DER uses collar 0 with overlap scored and is the total error time over the total speech time of the set; JER is the per-recording mean; all counts are totals over the set.

\begin{table}[t]
\centering
\caption{Sets used for counting. Speakers are reference speakers.}
\label{tab:sets}
\small
\begin{tabular}{lrr}
\toprule
Set & Recordings & Speakers \\
\midrule
All (Sec.~5) & 990 & 4{,}362 \\
\quad $\leq 8$ / $\geq 9$ speakers & 909 / 81 & 3{,}468 / 894 \\
eval (Sec.~7) & 681 & 2{,}954 \\
\quad $\leq 8$ / $\geq 9$ speakers & 623 / 58 & 2{,}307 / 647 \\
calib MSDWild (Sec.~7.3) & 59 & 319 \\
\bottomrule
\end{tabular}
\end{table}

\textbf{Systems.}
\emph{Nemotron-3} (nvidia/Nemotron-3-Diarization)~\cite{nemotron3} is a streaming Sortformer~\cite{sortformer,streaming-sortformer} with eight output slots; we run it once per recording with NeMo 3.1.0 in a high-latency setting (30.4\,s input buffer) and threshold each slot at 0.5.
\emph{fast} is the fast pyannote-3.1-style configuration of the voxmap diarization library\footnote{\url{https://github.com/panchorange/voxmap}}: segmentation-3.0~\cite{powerset} local speakers clustered agglomeratively with CAM++~\cite{campplus} embeddings, with no cap on the number of speakers.

\textbf{Environment.}
Everything runs on one MacBook Pro (Apple M4, 24\,GB).
Nemotron-3 and fast run on the Apple GPU (MPS), except four Nemotron-3 recordings longer than about 87 minutes, which run on CPU\footnote{Recordings longer than about 65{,}536 frames of 80\,ms produce almost no output on MPS. Computing only the input features on CPU lets the rest run on MPS.}.
For the first pass of global assignment, these four run with features on CPU and the rest on MPS. Speaker embeddings (CAM++) are computed on CPU.
RTFx is total audio duration over total inference time for the set, excluding audio and model loading.

\section{Analysis of Lost Speakers}
\label{sec:analysis}
\begin{figure*}[t]
\centering
\begin{minipage}{0.35\textwidth}\centering\includegraphics[width=\linewidth]{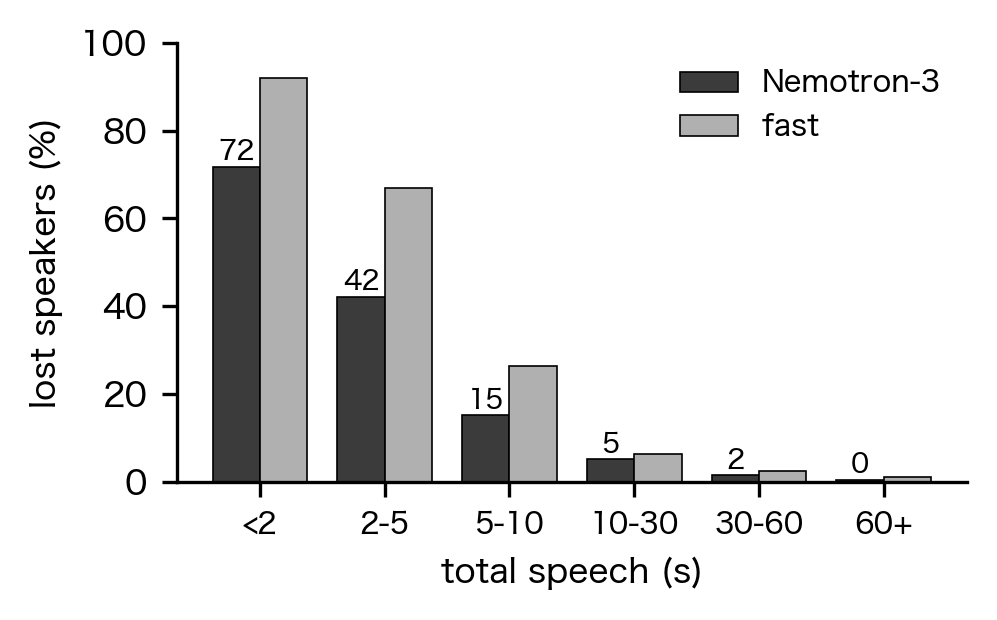}\end{minipage}\hfill
\begin{minipage}{0.35\textwidth}\centering\includegraphics[width=\linewidth]{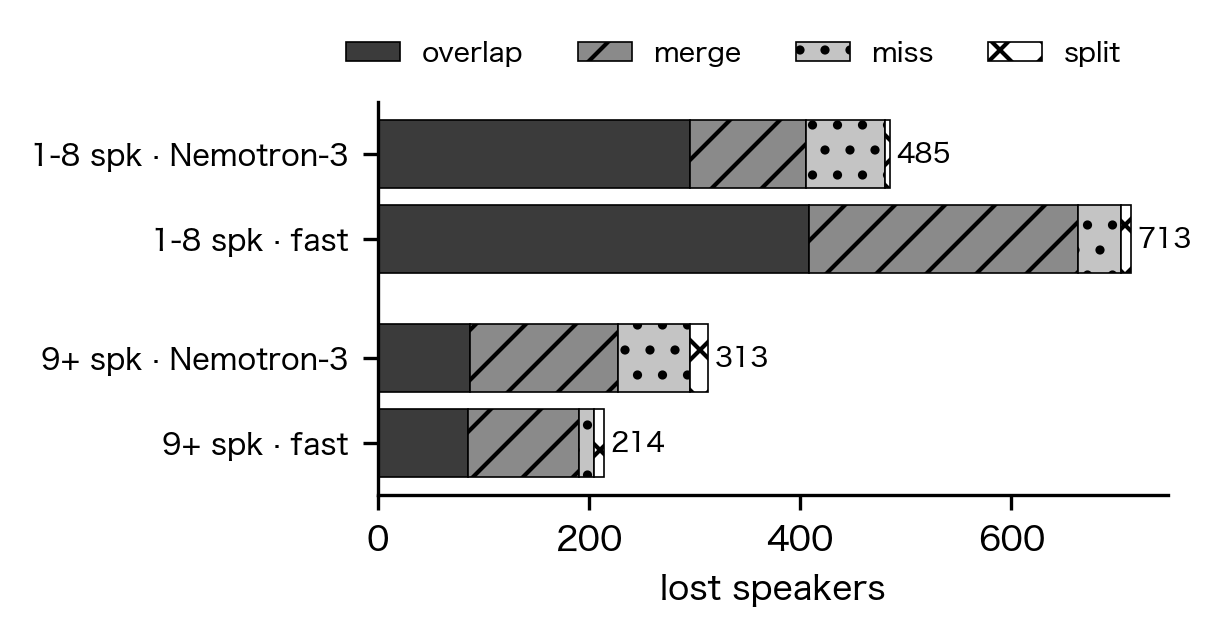}\end{minipage}
\caption{Left: fraction of lost speakers by total speech (all, up to eight speakers, 909 recordings); the denominator is the number of reference speakers in each bin, numbers are Nemotron-3.
Right: lost speakers by type and number of speakers (all 990 recordings). With nine or more speakers, Nemotron-3 has more merges and misses than fast.}
\label{fig:analysis}
\end{figure*}

\textbf{Who is lost.}
In the recordings with up to eight speakers (909 recordings, 3{,}468 reference speakers), Nemotron-3 loses 485 speakers (14.0\%), and at least one speaker is lost in 31\% of the recordings. fast loses 713 (20.6\%).
Loss depends strongly on speaking time (Fig.~\ref{fig:analysis}, left): Nemotron-3 loses 181 of the 252 speakers with less than 2\,s of speech (72\%), but only 4 of the 921 speakers with 60\,s or more (0.4\%).
Of the 485 lost speakers, 425 (88\%) are low-activity speakers; for fast, 633 of 713 (89\%).
Some are non-verbal: for 411 of the 1{,}467 low-activity speakers in VoxConverse and MSDWild (28\%), the reference segments are not speech-like per the AST audio event classifier~\cite{ast} (voices under music, singing, laughter, coughs); we count all speakers as annotated.

\textbf{How speakers are lost.}
The type depends on the number of speakers (Fig.~\ref{fig:analysis}, right).
With up to eight speakers, 296 of the 485 speakers Nemotron-3 loses (61\%) are overlaps.
With nine or more speakers (81 recordings, 894 speakers), 313 (35\%) are lost, and 209 of them (67\%) are merges or misses.
fast, which has no speaker cap, loses more than Nemotron-3 with up to eight speakers (713 vs.\ 485) but fewer with nine or more (214 vs.\ 313).
Indeed, Nemotron-3 outputs eight speakers in 67 of the 81 recordings with nine or more speakers (32 of 909 with up to eight).
We therefore attribute the merges and misses with nine or more speakers to the eight slots not fitting everyone, and treat the causes as \emph{slot shortage} (nine or more) and \emph{overlap} (up to eight).

\section{Remedies by Cause}
\label{sec:method}
None of the methods retrains the model.

\begin{figure*}[t]
\centering
\includegraphics[width=0.46\textwidth]{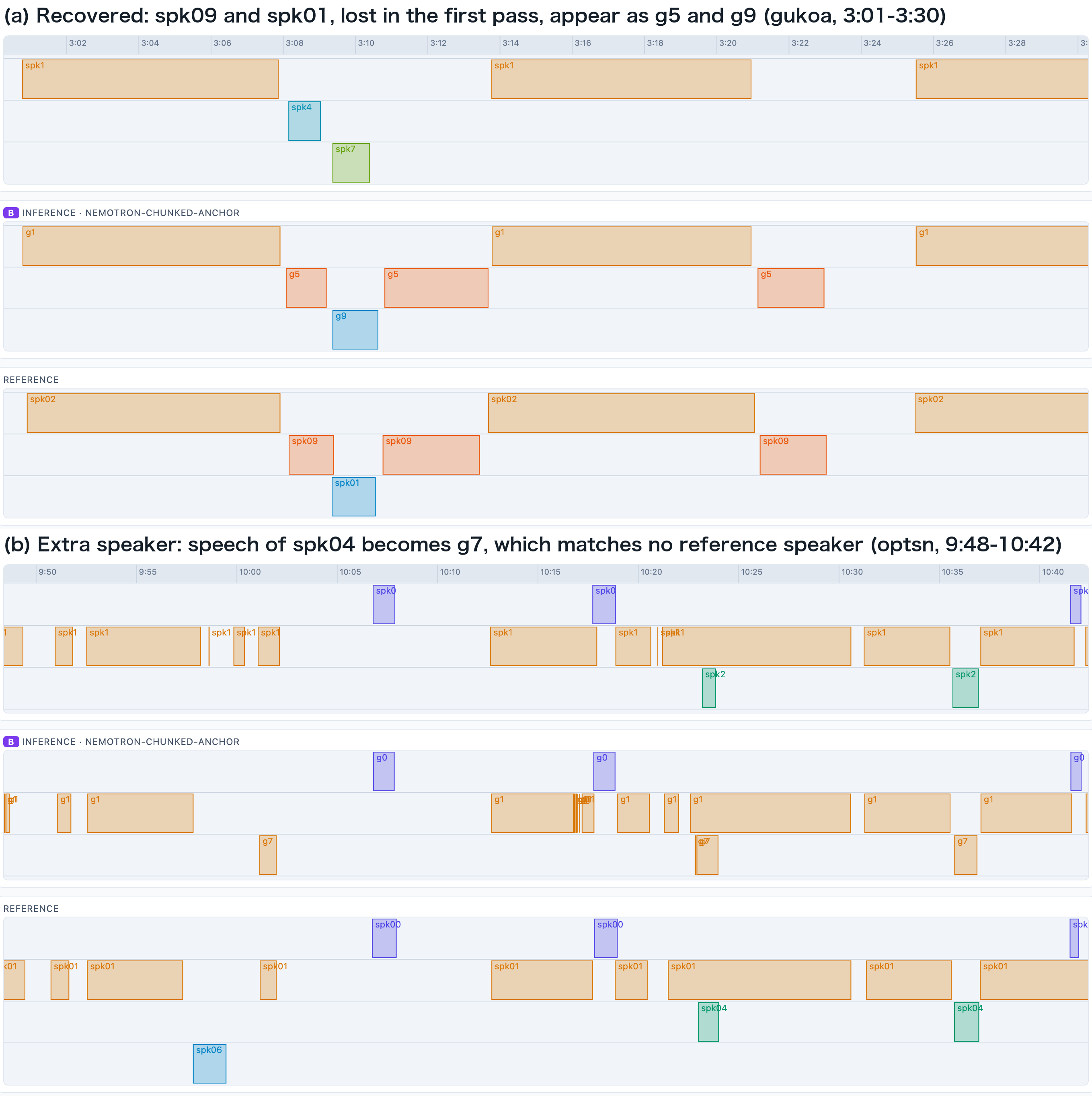}
\caption{Examples of global assignment (VoxConverse gukoa and optsn, voxmap studio). Each panel shows, from top, the Nemotron-3 output, the global-assignment output (labelled with its experiment name, chunked-anchor) and the reference; rows of speakers with no segments are omitted.}
\label{fig:results}
\end{figure*}

\begin{table*}[t]
\centering
\caption{Results on eval (681 recordings). DER: collar 0, overlap scored (in parentheses: collar 0.25, overlap excluded).
Lost, detected and extra speakers are totals over the 681 recordings; the lost breakdown is overlap / merge / miss / split, and detected counts out of 999 low-activity speakers.
RTFx is over the 681 recordings; global assignment and adaptive include the first-pass inference used for gating.}
\label{tab:main}
\small
\begin{tabular}{lrrrrrr}
\toprule
System & DER (\%) & JER (\%) & Lost & Detected & Extra & RTFx \\
\midrule
fast & 23.11 (12.82) & 45.6 & 629 (312 / 265 / 37 / 15) & 314 & 397 & 193 \\
Nemotron-3 (baseline) & 16.56 (9.39) & 29.7 & 546 (251 / 168 / 111 / 16) & 516 & 50 & 96 \\
Adaptive & 15.67 (8.14) & 27.9 & 396 (243 / 80 / 66 / 7) & 577 & 236 & 71 \\
Global assignment (ours) & 15.57 (8.16) & 28.2 & 417 (246 / 97 / 68 / 6) & 560 & 100 & 85 \\
\bottomrule
\end{tabular}
\end{table*}

\subsection{Slot shortage: windowed re-inference with global assignment}
\textbf{Gate.} Nemotron-3 is first run on the whole recording; we continue only if the output reaches eight speakers and the recording is shorter than 30 minutes.
Of the 309 calib recordings, 18 of the 31 whose first-pass output had eight speakers had nine or more reference speakers, versus 5 of the 278 with seven or fewer.
\textbf{Windows.} The audio is cut into 120\,s windows and Nemotron-3 is run per window; within a window, even crowded recordings fit in eight speakers.
\textbf{Linking.} For each window speaker we collect up to 20\,s of their solo output, longest first, extract one CAM++ embedding, and repeatedly merge the pair with the highest centroid cosine similarity while it is at least 0.6; two speakers from the same window are never merged.
\textbf{Global assignment.} Window speakers with less than 10\,s of speech have unstable embeddings, fail to link and tend to become extra speakers. They are held out of linking and finally assigned to the group of the first-pass (global) speaker that overlaps at least half of their speech (kept as separate speakers if the embedding similarity is below 0.2). We call this \emph{global assignment}.
All thresholds were chosen on calib.
For comparison we also report an \emph{adaptive} variant that neither gates on duration nor uses global assignment (600\,s windows for recordings of 30 minutes or more).

\subsection{Overlap: adding a hidden second speaker}
\label{sec:overlap-method}
We tried four methods.
(i) \emph{Probabilities}: following overlap-aware reassignment~\cite{bullock}, in frames where segmentation-3.0 estimates more concurrent speakers than are output, add Nemotron-3 slots below 0.5 in order of probability.
(ii) \emph{Embeddings}: compare the frame's embedding with speakers absent from the window, add the most similar one, or create a new speaker if none is similar.
(iii) \emph{Separated embeddings}: split the candidate audio into two sources with MossFormer2~\cite{mossformer2} and use the embedding of the source unlike the active speaker.
(iv) \emph{PixIT}: run PixIT (pyannote/speech-separation-ami-1.0) separately and add PixIT speakers that match no speaker of the global-assignment output.

\section{Results and Discussion}
\label{sec:results}
Table~\ref{tab:main} shows overall eval results and Table~\ref{tab:bands} splits them by the number of speakers.

\begin{table}[t]
\centering
\caption{eval results by number of speakers. DER: collar 0, overlap scored. Detected counts out of the low-activity speakers (790 with up to eight, 209 with nine or more). p: one-sided Wilcoxon test for an increase in extra speakers over the baseline.}
\label{tab:bands}
\small
\setlength{\tabcolsep}{3.5pt}
\begin{tabular}{llrrrr}
\toprule
Set & System & DER & Lost & Det. & Extra (p) \\
\midrule
\multirow{3}{*}{\shortstack[l]{$\leq 8$ spk.\\623 rec.}} & Baseline & 15.74 & 318 & 448 & 45 \\
 & Global & 15.71 & 316 & 449 & 50 (0.13) \\
 & Adaptive & 15.82 & 317 & 449 & 104 ($<$0.001) \\
\midrule
\multirow{3}{*}{\shortstack[l]{$\geq 9$ spk.\\58 rec.}} & Baseline & 20.10 & 228 & 68 & 5 \\
 & Global & 14.99 & 101 & 111 & 50 ($<$0.001) \\
 & Adaptive & 14.99 & 79 & 128 & 132 ($<$0.001) \\
\bottomrule
\end{tabular}
\end{table}

\subsection{Slot shortage is halved without training}
On the eval recordings with nine or more speakers (58), lost speakers drop from 228 (baseline) to 101 (global assignment), and DER from 20.10\% to 14.99\% (losses drop in 36 of 58 recordings; one-sided Wilcoxon test on per-recording differences, $p < 10^{-7}$).
The reduction comes mainly from merges (103 to 33) and misses (58 to 15), the types attributed to slot shortage in Sec.~\ref{sec:analysis}, and detected low-activity speakers rise from 68 to 111 out of 209.
In Fig.~\ref{fig:results}(a), spk09 and spk01, lost in the first pass, appear as their own speakers g5 and g9.

On the same 58 recordings, however, extra speakers grow from 5 to 50 (0.8 per recording on average; 24 recordings changed, $p < 0.001$).
In Fig.~\ref{fig:results}(b), speech of spk04 that the first pass matched correctly fails to link across windows after windowing and becomes g7, which matches no reference speaker.
The adaptive variant reduces losses further to 79 on these 58 recordings but yields 132 extra speakers.
Global assignment yields fewer extra speakers because it returns short fragments to first-pass speakers, though it can also return a different person the first pass had merged; since it barely changes recordings with up to eight speakers, we take it as our main method.
Windowing costs inference: eval RTFx is 96 (baseline), 85 (global assignment) and 71 (adaptive).

\subsection{Recordings with up to eight speakers do not change}
On the eval recordings with up to eight speakers (623), lost speakers are 318 (baseline) vs.\ 316 (global assignment) and DER 15.74\% vs.\ 15.71\%, almost unchanged.
Per-speaker recall and the number of output speakers are identical to the baseline in 612 of the 623 recordings.
Extra speakers total 45 (baseline) vs.\ 50 (global assignment). Only four recordings changed their count (differences $-1$, $+1$, $+1$, $+4$); with four recordings the test cannot go below $p=0.0625$ even if all increase, so $p=0.13$ says neither that extra speakers increased nor that they did not.
Global assignment thus neither harms nor improves these recordings. The adaptive variant changes the count in 15 of the 623 recordings and raises extra speakers to 104 ($p < 0.001$).

Of the 316 speakers still lost there, 195 are overlaps; for the remaining 64 merges and 53 misses, which occur although slots are available, we found no remedy and leave them unresolved.

\subsection{Overlap is not recovered without training}
\label{sec:overlap}
None of the four methods substantially reduced overlap losses.
(i) reduced overlap losses on eval (681) by 15\%, but merge losses increased by 39\% and DER (collar 0.25, overlap excluded) rose from 13.50\% to 14.73\%\footnote{Only (i) is compared with a baseline that ran the four long recordings on MPS.}.
(ii), applied on top of an output that windows every recording at 120\,s, reduced overlap losses on eval from 251 to 185, but the recovered speakers had a median recall of 0.16 (fragments of speech), and extra speakers grew from 50 to 1{,}968.

(iii) and (iv) were piloted on the 59 calib MSDWild recordings in which Nemotron-3 has overlap losses (118 speakers lost to overlap; Table~\ref{tab:overlap}).
(iii) left the hit rate of added speakers unchanged relative to mixed audio.
(iv) gave nonzero recall to 62 of the 118, but the added speakers paint long stretches of the recording (32 of 33\,s in one case) and hit hidden speakers only by chance.
Counting only those whose recall is at least three times the fraction of the recording the speaker was output for (\emph{well-placed}), just 13 of 118 remain (8 detected), and DER on the 59 recordings rises from 19.98\% (global assignment) to 48.91\%.

\begin{table}[t]
\centering
\caption{Pilot of overlap remedies (59 calib MSDWild recordings). DER: collar 0.25, overlap excluded. Recovered ($\mathrm{recall}>0$), detected and well-placed count out of the 118 speakers lost to overlap; extra is the total over 59 recordings. --: not computed. The well-placed threshold (3$\times$) was set after seeing (iv).}
\label{tab:overlap}
\small
\setlength{\tabcolsep}{3pt}
\begin{tabular}{lrrrr}
\toprule
System & DER & Rec. / det. & Well-placed & Extra \\
\midrule
Global assignment & 19.98 & 0 / 0 & 0 / 0 & 4 \\
+ (ii) embeddings & 21.64 & 19 / 2 & -- & 28 \\
+ (iii) separated & 21.68 & 21 / 2 & -- & 28 \\
+ (iv) PixIT speakers & 48.91 & 62 / 33 & 13 / 8 & 35 \\
\bottomrule
\end{tabular}
\end{table}

\section{Conclusion}
In an eight-slot end-to-end diarizer, nine in ten lost speakers speak for less than 10\,s.
Of the two causes, only slot shortage (nine or more speakers) could be fixed without training; with up to eight speakers, neither overlaps (60\% of losses) nor the remaining merges and misses were reduced by the training-free methods we tried.
Recovering short speech hidden under others will likely require training, such as speaker-weighted losses~\cite{beyond-der} or synthetic overlap data, or methods that model overlap directly such as target-speaker voice activity detection~\cite{tsvad}.
Limitations: no comparison with recent systems such as DiariZen~\cite{diarizen}; non-official calib/eval splits, with the 30-minute condition also chosen after seeing eval results; loss types are our own definition, not verified by listening.


\begin{thebibliography}{99}\scriptsize
\setlength{\itemsep}{0pt}\setlength{\parskip}{0pt}
\bibitem{sortformer} T.~Park \emph{et al.}, ``Sortformer: A novel approach for permutation-resolved speaker supervision in speech-to-text systems,'' in \emph{Proc. ICML}, 2025.
\bibitem{nemotron3} I.~Medennikov \emph{et al.}, ``Nemotron 3 Diarization: Advancing the accuracy-efficiency frontier in streaming speaker diarization,'' arXiv:2611.12246, 2026.
\bibitem{streaming-sortformer} I.~Medennikov \emph{et al.}, ``Streaming Sortformer: Speaker cache-based online speaker diarization with arrival-time ordering,'' in \emph{Proc. Interspeech}, 2025.
\bibitem{dihard} N.~Ryant \emph{et al.}, ``The second DIHARD diarization challenge: Dataset, task, and baselines,'' in \emph{Proc. Interspeech}, 2019.
\bibitem{ber} T.~Liu and K.~Yu, ``BER: Balanced error rate for speaker diarization,'' arXiv:2211.04304, 2022.
\bibitem{cder} G.~Cheng \emph{et al.}, ``The conversational short-phrase speaker diarization (CSSD) task: Dataset, evaluation metric and baselines,'' arXiv:2208.08042, 2022.
\bibitem{beyond-der} L.~Samarakoon and H.~Zhu, ``Beyond DER: Speaker counting in crowded end-to-end diarization,'' in \emph{Proc. SLT}, 2026.
\bibitem{eendeda} S.~Horiguchi, Y.~Fujita, S.~Watanabe, Y.~Xue, and P.~Garc\'{i}a, ``Encoder-decoder based attractors for end-to-end neural diarization,'' \emph{IEEE/ACM Trans. Audio, Speech, Lang. Process.}, 2022.
\bibitem{eendvc} K.~Kinoshita, M.~Delcroix, and N.~Tawara, ``Integrating end-to-end neural and clustering-based diarization: Getting the best of both worlds,'' in \emph{Proc. ICASSP}, 2021.
\bibitem{diarizen} J.~Han, F.~Landini, J.~Rohdin, A.~Silnova, M.~Diez, and L.~Burget, ``Leveraging self-supervised learning for speaker diarization,'' in \emph{Proc. ICASSP}, 2025.
\bibitem{eendgla} S.~Horiguchi, S.~Watanabe, P.~Garc\'{i}a, Y.~Takashima, and Y.~Kawaguchi, ``Online neural diarization of unlimited numbers of speakers using global and local attractors,'' \emph{IEEE/ACM Trans. Audio, Speech, Lang. Process.}, vol.~31, pp.~706--720, 2023.
\bibitem{chen-sortformer} Y.-W.~Chen \emph{et al.}, ``From who said what to who they are: Modular training-free identity-aware LLM refinement of speaker diarization,'' arXiv:2509.15082, 2025.
\bibitem{pixit} J.~Kalda \emph{et al.}, ``PixIT: Joint training of speaker diarization and speech separation from real-world multi-speaker recordings,'' in \emph{Proc. Odyssey}, 2024.
\bibitem{icsi} A.~Janin \emph{et al.}, ``The ICSI meeting corpus,'' in \emph{Proc. ICASSP}, 2003.
\bibitem{ami} J.~Carletta \emph{et al.}, ``The AMI meeting corpus: A pre-announcement,'' in \emph{Proc. MLMI}, 2005.
\bibitem{voxconverse} J.~S.~Chung \emph{et al.}, ``Spot the conversation: Speaker diarisation in the wild,'' in \emph{Proc. Interspeech}, 2020.
\bibitem{msdwild} T.~Liu \emph{et al.}, ``MSDWild: Multi-modal speaker diarization dataset in the wild,'' in \emph{Proc. Interspeech}, 2022.
\bibitem{powerset} A.~Plaquet and H.~Bredin, ``Powerset multi-class cross entropy loss for neural speaker diarization,'' in \emph{Proc. Interspeech}, 2023.
\bibitem{campplus} H.~Wang \emph{et al.}, ``CAM++: A fast and efficient network for speaker verification using context-aware masking,'' in \emph{Proc. Interspeech}, 2023.
\bibitem{bullock} L.~Bullock, H.~Bredin, and L.~P.~Garcia-Perera, ``Overlap-aware diarization: Resegmentation using neural end-to-end overlapped speech detection,'' in \emph{Proc. ICASSP}, 2020.
\bibitem{ast} Y.~Gong, Y.-A.~Chung, and J.~Glass, ``AST: Audio spectrogram transformer,'' in \emph{Proc. Interspeech}, 2021.
\bibitem{tsvad} I.~Medennikov \emph{et al.}, ``Target-speaker voice activity detection: A novel approach for multi-speaker diarization in a dinner party scenario,'' in \emph{Proc. Interspeech}, 2020.
\bibitem{mossformer2} S.~Zhao \emph{et al.}, ``MossFormer2: Combining transformer and RNN-free recurrent network for enhanced time-domain monaural speech separation,'' in \emph{Proc. ICASSP}, 2024.
\end{thebibliography}
\end{document}